# Integrating psychotherapy practices and gamified elements in novel game mechanics for stress relief

Styliani Zygotegou[1], Georgios Anastassakis[1], Georgios Tsatiris[1,2], Kostas Karpouzis[3]

[1] *Department of Games, SAE Athens, Greece*
[2] *Artificial Intelligence and Learning Systems Laboratory, National Technical University of Athens, Greece*
[3] *Dept. of Communication, Media and Culture, Panteion University of Social and Political Science, Athens, Greece*

**Abstract**
We explore novel game mechanics and techniques in the domain of gamified and game-based mobile mental health applications. By combining modern game design elements with techniques applied by practitioners (e.g., therapists) and known mechanics used in relevant games, we developed an integrated mobile game. Playtesting with a group of individuals showed a positive response towards the study's claims and a promising direction for further research.

## 1. Introduction

Today's fast-paced way of life entails a great deal of stress-inducing situations. The video game industry has expanded its interests towards psychology-related applications in recent years. In contrast to arguments against the use of games for purposes other than having fun, new studies have emerged, aiming to prove that casual gaming can act as a stress relief factor. Many practitioners in the field of psychology and psychotherapy have collaborated with programmers and game designers and implemented games and gamified applications for treating certain cases of mental illness. These applications come under the broad term of serious games, which include those specially designed games aiming at tackling deeper issues than providing entertainment (i.e., rehabilitation, learning, etc.).

Psychotherapy is applied using a vast arsenal of techniques and established methods. Most of them have some concepts in common, such as the need to help the patient feel comfortable and safe and facilitate their attempt to get in touch with the source of their stress. This study aims to prove the direct connection between established methods of psychotherapy and the design of relevant applications and games. Practitioners have been interviewed and consulted during the design and implementation processes, to ensure that the final product will meet the necessary requirements.

In this paper, we describe the design, implementation, and early evaluation of a gamified mobile application for stress control. The design process was based on feedback from experts and literature in related fields, resulting in a selection of requirements and functionalities, which the mobile app should respect; this process is described in Section 2. Following that, we describe the application as it's being used by users and qualitative input from experts and users.

## 2. Game design

Stress can be induced by a variety of factors that affect our everyday life; these range from our socioeconomic status, our general health, the assumption of risks and initiatives, all the way to our choices of people we hang around or means to have fun. In this context, daily mental activities may serve as a valuable protection measure [16]. Other usual components affecting stress, include globalization and new technological regimes [17], and career progress and stability (or lack thereof). Especially in the latter case, stress may result in *rumination*, a mental state in which people go over the same thoughts over and over during the day, resulting in them being social isolated, neglecting everyday tasks and chores, and underperforming in their work-related assignments. This state also has adverse effects with respect to a person's general health, since it is commonly associated with elevated heart rate, cortisol levels, and blood pressure [18].

According to directives from the National Institute of Mental Health, since the effects of continuous and untreated stress tend to increase over time, it should be treated using tools or activities for stress management. These include:
- Acknowledging somatic response to stress, such as having trouble sleeping, increased use of alcohol or other substances, increased outbursts of irritation, feelings of depression and decreased energy levels
- Consultation with a therapist and a general practitioner (to take into account pre-existing health problems or other issues)
- Frequent physical exercise: even a daily walk lasting around 30 minutes can assist in reducing stress and improve one's mood
- Relaxing mental exercises, such as meditation, yoga, or tai chi
- Setting measurable and realistic goals and expectations: prioritization of tasks to be completed is very important here, as well as making sure one does not assume new tasks which may lead to overloading
- Social interaction with people who can offer emotional support

In our work, we chose to implement calming tasks for the players, taking into account the Experiential Avoidance Model [19], which suggests that people in extreme and long-lasting stress may opt to inflict harm to themselves, turning the mental anguish to physical, which can be coped with more easily. In order to eliminate the option of Deliberate Self-Harm (often abbreviated as DSH) as a coping mechanism, harming activities must be replaced by those which promote the person's well-being and do not cause any damage. Ideally, these activities should keep both the person's mind and body occupied, resulting in video games being an obvious choice for the task.

The above concepts have been put to use to design the activities in a mobile game, called 'Embrace' implemented using the Unity game engine (see Figure 1 for details on the game's functionalities). The game consists of four levels, with increasing difficulty, so as to keep the players' interest at all times and offer them a sense of progression which is essential to create and maintain flow [20]. The aim of the players in each level is to guide a ball towards a goal, avoiding obstacles which may hinder their attempts. When the level is completed successfully, a dialog window is unlocked, and players are given access to the next level. Important therapy steps implemented as scenarios in the game include the process of *grounding* the patient to the present, the sense of *security*, the *projection* of the stressful feeling and the contact with it as well as its acceptance and the process of *stripping* it from any logical causes or foundation.

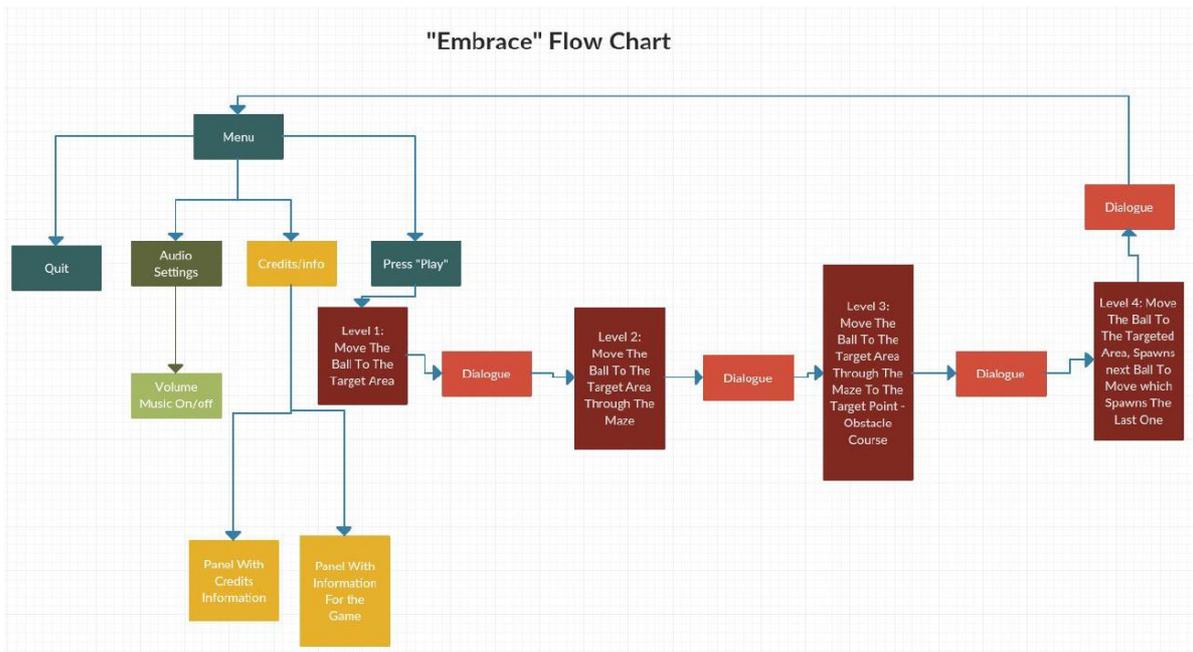

**Figure 1:** A flow chart indicating gameplay in Embrace

In the first level, players must guide the ball towards the ground. In general, the ball represents the player's feelings, something unconsciously prevalent, which still is made known to the player during the game. The idea behind this scenario is to help the player feel grounded in the present. In the second level, the player needs to guide the ball through a relatively easy maze, with openings between vertically placed platforms. This gameplay helps players establish a connection with the ball, which acts as their projection in the game world. The third level consists of another maze, this time with moving obstacles, which represent logical fallacies and hypothetical thinking which can hamper the progress of the patient at the beginning of the therapeutic process, leading to extended rumination. Finally, the fourth level focuses on acceptance: here, players become acquainted with the feelings of joy, fear and stress. Between levels, descriptive dialog messages are shown to players, aiming to guide them both through the therapeutic process and through each level of gameplay.

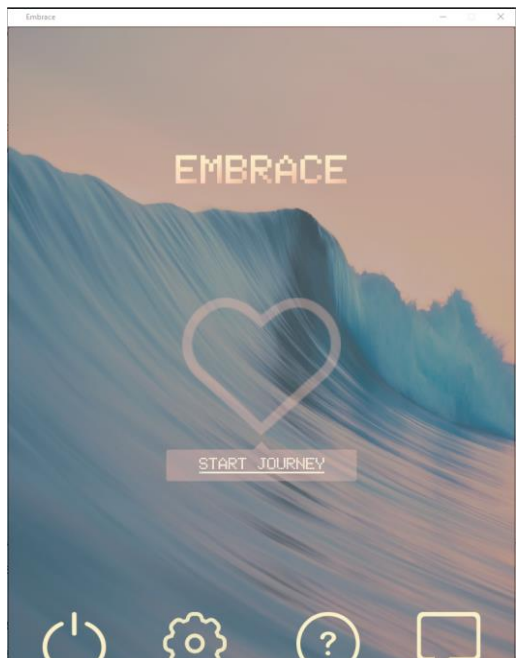
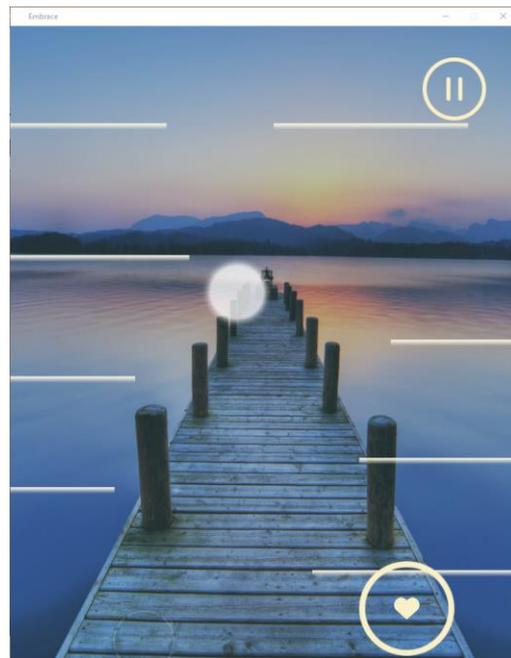

**Figure 2a:** 'Embrace' opening screen    **Figure 2b:** Design and gameplay of level 1

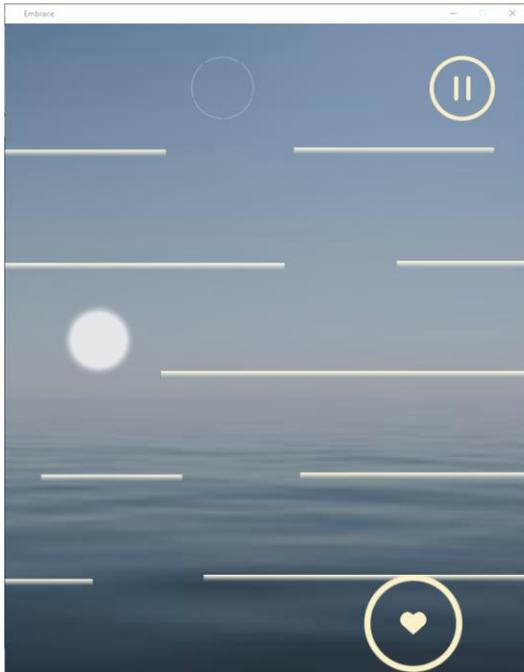 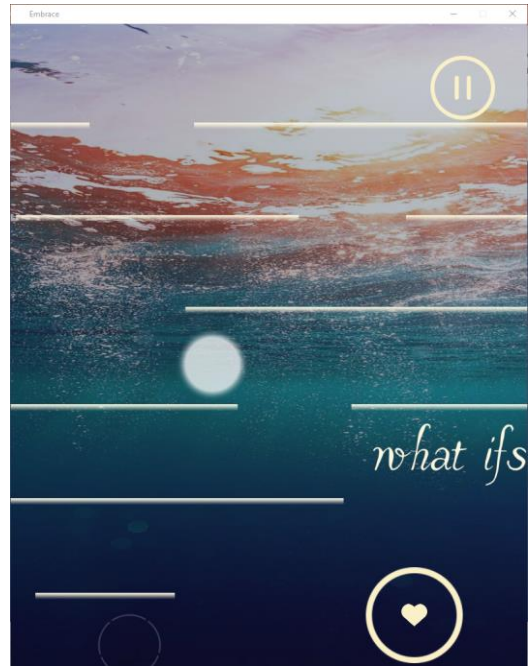

**Figure 2c:** Design and gameplay of a harder level    **Figure 2d:** Design and gameplay of the final level

All interfaces follow pastel-like color patterns as the use of light, warm colors with tones of white have been shown to help users relax. Furthermore, circular shapes and curves have been promoted for the interface design as the connection of feelings and preference of users towards shapes without many corners have been studied and verified. This design choice also follows the Eleven Principles of the Gestalt method, in which symmetry, harmony, consistency and simplicity are preferable to the neutral user.

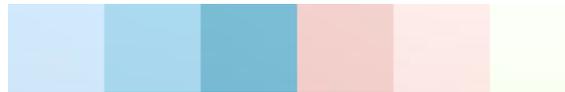

**Figure 3:** Embrace user interface and game palette

## 3. Experimentation and conclusions

The application was playtested by a small group of five individuals between 20 and 27 years of age. Before playing, the subjects were asked to sit comfortably and keep their feet on the ground. Users found these preconditions to be important for the effectiveness of the application.

Results showed positive reactions towards the goal of the application. Two of the five subjects reported feeling calmer after playing and four of them reported general positive feelings. The focus of each level was apparent to the subjects and those that were not fully immersed, were able to understand its meaning and provide thoughtful comments. This proves that the concepts and the design principles integrated in this application are in the right way and can serve as a viable start for more sophisticated gamified or serious game applications.

## Author biographies

Styliani (Stella) Zygotegou received her BSc in Game Programming from the University of Middlesex. She works as a game and graphic designer and game developer.

George Anastassakis is a Senior Lecturer in Games Programming at SAE Athens. He received his PhD in Informatics from the Department of Informatics of the University of Piraeus in 2010. The subject of his thesis was REVE, an abstract representation for virtual environments. In 2000, he received his MSc in Multimedia Technology from the Department of Computation, UMIST (University of Manchester Institute of Science and Technology). His MSc dissertation, titled "Intelligent Agents in Virtual Worlds", introduced a simple intelligent agent system named VITAL, which was later advanced to a multi-agent level.

George Tsatiris is a PhD candidate at the National Technical University of Athens and Lecturer at the SAE Athens Department of Game Programming. His research interests include image and video analysis and understanding, affective computing and serious games.

Kostas Karpouzis is an assistant professor at the Department of Communication, Media and Culture, Panteion University of Social and Political Sciences, in Athens, Greece. In his research, he's looking for ways to make computer systems more aware of and responsive to the way people interact with each other, using natural, affective interaction and intelligent, user-adaptive algorithms. Kostas is also investigating how gamification and digital games can be used in classroom and informal settings to assist conventional teaching and help teach social issues and STEAM subjects to children and adults. Two of the games he has worked on have won the Best Serious Game award from the Games and Learning Alliance: Siren (2013) and Navigo (2018).